# REFINING DATA SECURITY IN INFRASTRUCTURE NETWORKS SUPPORT OF MULTIPATH ROUTING

K. Karnavel[1], L. shalini[2], M. Ramananthini[3]

[1]Assistant Professor, [2,3]PG Student, Department of Computer Science & Engineering, AIHT, Chennai
treseofkarnavel@gmail.com

**Abstract**
*An infrastructure network is a self-organizing network with help of Access Point (AP) of wireless links connecting nodes to another. The nodes can communicate without an ad hoc. They form an uninformed topology (BSS/ESS), where the nodes play the role of routers and are free to move randomly. Infrastructure networks proved their efficiency being used in different fields but they are highly vulnerable to security attacks and dealing with this is one of the main challenges of these networks at present. In recent times some clarification are proposed to provide authentication, confidentiality, availability, secure routing and intrusion avoidance in infrastructure networks. Implementing security in such dynamically changing networks is a hard task. Infrastructure network characteristics should be taken into consideration to be clever to design efficient solutions. Here we spotlight on civilizing the flow transmission privacy in infrastructure networks based on multipath routing. Certainly, we take benefit of the being of multiple paths between nodes in an infrastructure network to increase the confidentiality robustness of transmitted data with the help of Access Point. In our approach the original message to secure is split into shares through access point that are encrypted and combined then transmitted along different disjointed existing paths between sender and receiver. Even if an intruder achieve something to get one or more transmitted distribute the likelihood that the unique message will be reconstituted is very squat.*

*Keywords* - *Infrastructure networks Security Confidentiality Multipath routing Basic Service Set (BSS) and Extended Service Set (ESS).*

---***---

## 1. INTRODUCTION

An 802.11 networking framework in which devices communicate with each other by first going through an Access Point (AP). In infrastructure mode, wireless devices can communicate with each other or can communicate with a wired network. When one AP is connected to wired network and a set of wireless stations it is referred to as a Basic Service Set (BSS). An Extended Service Set (ESS) is a set of two or more BSSs that form a single sub network. These base stations provide access for mobile terminals to a backbone wired network. Network control functions are performed by the base stations, and often the base stations are connected together to facilitate coordinated control. Base station coordination in infrastructure-based networks provides a centralized control mechanism for transmission scheduling, dynamic resource allocation, power control, and handoff. Handoff management has widely been recognized as one of the most important and challenging problems for a seamless access to wireless network. Most corporate wireless LANs operate in infrastructure mode because they require access to the wired LAN in order to use services such as file servers or printers. compared to ad-hoc networks it efficiently utilizes network resources and results in high data rates and lower delays and loss. It also offers the advantage of scalability, centralized security management and improved reach. Flow transmission confidentiality is improved in infrastructure through multipath routing. Multipath routing is the routing technique of using multiple alternative paths through a network, which can yield a variety of benefits such as fault tolerance, increased bandwidth, or improved security. The multiple paths computed might be overlapped, edge-disjoint or node-disjoint with each other. Extensive research has been done on multipath routing techniques, but multipath routing is not yet widely deployed in practice.

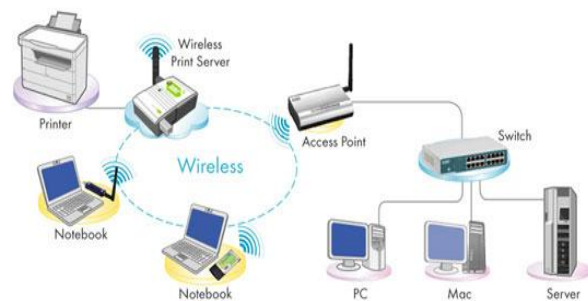

**Fig: 1** Infrastructure Networks based on multipath Routing

Based on that Fig.1 to improve performance of data sharing in the network and also refining the data security for transmitting the data from one node to another node. Basic service set operation provided signals through access point from one infrastructure to other. In BSS, provide high data security by support of access point in the wireless sensor networks for multipath routing.

CMR (Concurrent Multipath Routing) is often taken to mean simultaneous management and utilization of multiple available paths for the transmission of streams of data emanating from an application or multiple applications. In this form, each stream is assigned a separate path, uniquely to the extent supported by the number of paths available. If there are more streams than available paths, some streams will share paths. This provides better utilization of available bandwidth by creating multiple active transmission queues. It also provides a measure of fault tolerance in that, should a path fail, only the traffic assigned to that path is affected, the other paths continuing to serve their stream flows; there is also, ideally, an alternative path immediately available upon which to continue or restart the interrupted stream.

This method provides better transmission performance and fault tolerance by providing:
- Simultaneous, parallel transport over multiple carriers.
- Load balancing over available assets.
- Avoidance of path discovery when re-assigning an interrupted stream.

## 2. SECURITY

Security in wireless networks generally ranges from errors and omissions to threats to personal privacy. However, the more immediate concerns for wireless communications are device theft, denial of service, malicious hackers, malicious code, theft of service, and industrial and foreign espionage. Attacks resulting from these threats, if successful, place an agency's systems—and, more importantly, its data—at risk. Ensuring confidentiality, integrity, authenticity, and availability are the prime objectives of all government security policies and practices. Information must be protected from unauthorized, unanticipated, or unintentional modification. Security requirements include the following:
**Authenticity -** A third party must be able to verify that the content of a message has not been changed in transit.
**Non-repudiation -** The origin or the receipt of a specific message must be verifiable by a third party.
**Accountability -** The actions of an entity must be traceable uniquely to that entity.

## 3. RELATED WORK

Recently, several researches interesting in infrastructure networks security aspects (like authentication, availability, secure routing, intrusion detection, etc) do exist. We can classify existing approaches into five principal categories:
- Trust Models
- Key Management Models
- Routing Protocols Security
- Intrusion Detection Systems
- Multipath protocols

We mention below, some important proposals concerning each category:

### 3.1. Distributed Trust Model

The idea in[1] is based on the concept of trust. It adopts a decentralized management approach, generalizes the notion of rust, reduces ambiguity by using explicit trust statement and makes easier the exchange of trust-related information via a Recommendation Protocol. The Recommendation Protocol is used to exchange trust information. Entities that are able to execute the Recommendation Protocol are called agents. Agents use trust categories to express trust towards other agents and store reputation records in their private databases to use them to generate recommendations to other agents. In this solution, memory requirements for storing reputations and the behavior of the Recommendation Protocol are issues that were not treated.

### 3.2. Key Management Using Piconet

A Bluetooth network is called a small network or Piconet. It has eight stations. One of which is called primary station and the rest of the stations are called secondaries. All secondary stations synchronize their clocks and hoping sequence with the primary. A piconet can have only one primary station and the communication between them is one-one or one-many. Although a piconet can have a maximum of seven secondaries an additional eight secondaries can be in parked state. A secondary is synchronized with the primary but they cannot take part in communication until it is moved from the parked state. [3]

### 3.3. Key Agreement Based Password

The work developed in [8] draws up the scenario of a group wishing to provide a secured session in a conference room without supporting any infrastructure. The approach describes that these is a Weak Password that the entire group will share (for example by writing it on a board). Then, each member contributes to create a part of the session key using the weak password. This secured session key makes it possible to establish a secured channel without any centralized trust or infrastructure. This solution is adapted, to the case of conferences and meetings, where the number of nodes is small. It is rather strong solution since it does not have a strong shared key. But this model is not sufficient for more complicated environments. If we consider a group of people who do not know each other and want to communicate confidentially, this model becomes invalid. Another problem emerges if nodes are located in various places because the distribution of the Weak Password will not be possible any more.

### 3.4. Secure Multicast Communication In Infrastructure Network

Multicast is the most suitable model for reducing the incurring network load, when traffic needs to be securely delivered from a single authorized sender to a large group of valid receivers. Provision of security or multicast sessions is realized through enciphering the session traffic with cryptographic keys. All multicast members must hold valid keys in order to be able to decrypt the received information.

The problem of distributing and updating the cryptographic keys to valid members (key management), adds storage, communication and computational overhead to the network management. Key updates are required either periodically or on-demand, to accommodate membership changes in multicast groups. Security should consume as minimal energy as possible in updating the keys. Authors in [7] present a cross-layer algorithm that considers the node transmission power (physical layer) and the multicast routing tree (network layer) in order to construct an energy efficient key distribution scheme (application layer). That means they introduce a cross-layer design approach for key management in wireless multicast, that distributes cryptographic keys in an energy-efficient way. This solution is targeted towards a very special case (key management for multicast in infrastructure networks ensuring energy efficiency) and is complicated and supposes that the power consumption of computations is significantly reduced due to advances in silicon technologies, which is not true for all infrastructure network devices.

### 3.5. Secure Routing Protocol For Mobile Infrastructure Networks

An important aspect of infrastructure network security is routing security. The Secure Routing Protocol (SRP) presented in [9,6,12,10] counters malicious behavior that targets the discovery of topological information. SRP provides correct routing information (factual, up-to-date, and authentic connectivity information regarding a pair of nodes that wish to communicate in a secure manner). SRP discovers one or more routes whose correctness can be verified. After verification, route request are propagated to the desired trusted destination. Route replies are returned strictly over the reversed route, as accumulated in the route request packet. There is an interaction of the protocol with the IP layer functionality. The reported path is the one placed in the reply packet by the destination, and the corresponding connectivity information is correct, since the reply was relayed along the reverse of the discovered route. They ensure that attackers cannot impersonate the destination and redirect data traffic, cannot respond with stale or corrupted routing information, are prevented from broadcasting forged control packets to prevent the later propagation of legitimate queries, and are unable to influence the topological knowledge of benign nodes.

### 3.6. Intrusion Detection

One of recent interesting aspects of security in wireless networks, especially infrastructure networks is intrusion detection. It concerns detecting inappropriate, incorrect, or anomalous activity in the network. In [11] authors examine the vulnerabilities of wireless networks and argue that intrusion detection is a very important element in the security architecture for mobile computing environment. They developed such an architecture and evaluated a key mechanism in this architecture, anomaly detection for mobile infrastructure network through simulation experiments. Intrusion prevention measures, such as encryption and authentication, can be used in infrastructure networks to reduce intrusion, but cannot eliminate them. For example, encryption and authentication cannot defend against compromised mobile nodes, which often carry the private keys. In their architecture [11], they suggest that intrusion detection and response systems should be both distributed and cooperative to suit the needs of mobile infrastructure networks. Also, every node participates in intrusion detection and response. Thus there are individual IDS (Intrusion Detection Systems) agents placed on each node. If anomaly is detected in the local data, neighboring IDS agents will cooperatively participate in global intrusion detection actions. They conclude that intrusion detection can compliment intrusion prevention techniques (such encryption, authentication, secure MAC, secure routing, etc.) to secure mobile computing environment.

### 3.7. Security Based Multipath Routing Protocols

The aspect in which we are interested is this last one: security based multipath routing protocols. Multipath routing allows the establishment of multiple paths between a single source and single destination node. It is typically proposed in order to increase the reliability of data transmission (i.e., fault tolerance) or to provide load balancing [8]. Multipath routing has been explored in several different contexts. Traditional circuit switched telephone networks used a type of multipath routing called alternate path routing. Multipath routing has also been addressed in data networks which are intended to support connection oriented service with QoS like in ATM. We can take advantage of multipath routing to improve infrastructure network security. Let us assume that we have a secret message, if we send it through a single path, the enemy can compromise it by compromising any one of the nodes along this path. However, if we divide it into multiple parts and send these multiple parts via multiple independent paths, then the enemy has to compromise all the pieces from all the paths to compromise the message. Since 2003, some approaches were developed in this direction. Up until now there are only four protocols dealing with the infrastructure security problem using multipath routing (including our approach): SDMP, SPREAD [11].

### 3.8 Spread

The Security Protocol for Reliable Data Delivery (SPREAD) scheme addresses data confidentiality and data availability in a hostile infrastructure environment. The confidentiality and availability of message exchanges is enhanced through multipath routing with the help of SPREAD scheme assumptions like link encryption with neighboring nodes. The SPREAD scheme proposes the use of a(T,N) threshold secret sharing algorithm [5] to divide messages into multiple pieces. The simulation results of SPREAD show that the message interception probability, for both passive and active attacks, rapidly decreases with an increase in the number of paths used to transmit the message thereby enhancing data confidentiality Fig:2.

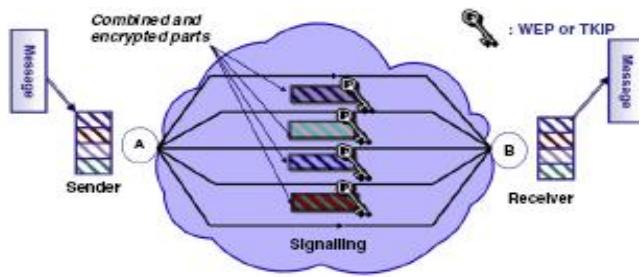

**Fig:2** Spreading the frame different path

## 4. SECURING DATA BASED MULTIPATH ROUTING IN INFRASTRUCTURE NETWORKS (SDMP)

The idea behind our protocol is to divide the initial message into parts then to encrypt and combine these parts by pairs. Then we exploit the characteristic of existence of multiple paths between nodes in an infrastructure networks to increase the robustness of confidentiality. This is achieved by sending encrypted combinations on the different existing paths between the sender and the receiver. In our solution, even if an attacker succeeds to have one part or more of transmitted parts, the probability that the original message can be reconstructed is low. We start by presenting our method principle.

### 4.1. Principle

At first, we present the principle of SDMP protocol in a simplified scheme, and then we explain in detail how it works.

The originality of our approach is that it does not modify the existing lower layer protocols. Some assumptions should be taken into consideration:

- The sender `A' and the receiver `B' are authenticated,
- WEP or TKIP is used for the encryption/decryption of frames at MAC layer and the authentication of the terminals,
- A mechanism of discovering the topology of the network is available,
- The protocol uses a routing protocol supporting multipath routing.

### 4.2 .Paths Selection In SDMP

In an infrastructure network to the topology for changes frequently depends on the transmission mode (BSS or ESS) from the wireless LAN or Wireless sensor networks, which makes wireless links instable. Sometimes packets might be dropped due to the bad wireless channel conditions, the collision at MAC layer transmission, or because of out of date routing information. When packet loss does occur, non-redundant share allocation will disable the reconstruction of the message at the intended destination. To deal with this problem, it is necessary to introduce some redundancy (if there is enough paths) in SDMP protocol to improve the reliability, (i.e. the destination would have better chance to receive enough shares for reconstructing the initial message). We propose that the decision of using or not redundancy will be taken according to the average mobility of the network's nodes and to the existing path number.

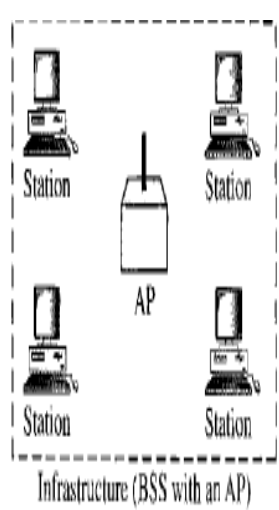 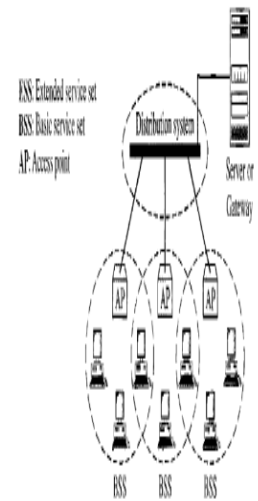

**Fig:3** Infrastructure network with BSS    **Fig:4** Infrastructure network with ESS

SDMP is based on multipath routing in infrastructure networks. The question is how to find the desired multiple paths in infrastructure network and how to deliver the different message parts to the destination using these paths? Routing in a infrastructure network presents great challenge because the nodes are capable of moving and the network topology(Fig.3 and Fig.4) can change continuously and unpredictably. The Dynamic Source Routing (DSR) protocol is capable of maintaining multiple paths from the source to a destination. This on-demand protocol works by broadcasting the route inquiry messages throughout the network and then gathering the replies from the destination.

## CONCLUSION

In this paper, we proposed a new solution that treats data confidentiality problem by exploiting a very important Infrastructure Network feature, which is used the existence of multiple paths between nodes. Our proposal improves data security efficiently without being time-consuming. It takes profit from existing Infrastructure Network characteristics and does not modify existing lower layer protocols. It is not complicated and can be implemented in different infrastructure network. This protocol is strongly based on multipath routing characteristics of infrastructure networks and uses a route selection based on security costs and BSS, ESS provide the high data security for secure data transmission via access point (AP). In this continuation our process before sending the data to check the path is idle or busy, afterword's, to send the data. The more the number of used paths is important, the more confidentiality is enforced. SDMP protocol can be combined with other solutions which consider other security aspects than confidentiality to improve significantly the efficiency of security systems in infrastructure network.